\documentclass[a4paper,11pt]{article}
\usepackage{authblk}
\pdfoutput=1
\usepackage[utf8]{inputenc}
\usepackage{pdflscape}
\usepackage{lscape}
\usepackage{array,booktabs,colortbl,multirow,tablefootnote}
\usepackage{makecell}
\usepackage{chngcntr}
\usepackage{relsize}
\usepackage{colortbl,colordvi,color}
\usepackage{subcaption}
\usepackage{todonotes}
\usepackage[toc,page]{appendix}

\usepackage{eurosym}
\usepackage{graphicx}
\usepackage{wrapfig,mdframed,caption}
\usepackage{comment}

\newcommand{\omegaL}{\omega_{L}}

\definecolor{cccc}{rgb}{0.4, 0.8, 0.3}

\newcommand{\ecrit}{{E}_\textrm{cr}}
\newcommand{\enLaser}{{\mathcal E}_\textrm{L}}

\begin{document}

\title{LUXE: A new experiment to study non-perturbative QED}
\author[a]{Louis Helary\thanks{louis.helary@desy.de \\~~~~~On behalf of the LUXE collaboration\\~~~~~Proceedings of the EPS-HEP 2021 conference\\}}
\affil[a]{Deutsches Elektronen-Synchrotron DESY\\ Notkestr. 85, 22607 Hamburg, Germany}

%
\maketitle
\section*{Abstract}

Laser Und XFEL Experiment (LUXE), is a new High Energy and Laser Physics experiment planned to be installed in DESY. It will study the interactions of a very-intense Laser beam with the high-energy, high-quality electron beams from the European XFEL accelerator, to measure characteristics of non-perturbative quantum electrodynamics.

\section{Introduction}
Quantum electrodynamics (QED) is a theory that explains the interaction between light and matter. It is one of the most well-tested physics theory; for instance the prediction of the electron anomalous moment (g-2) has a precision better than one part in a trillion, and the prediction agrees with measurements up to a very high level of accuracy~\cite{Hanneke_2011}.

The predictive power of QED mostly rely on the perturbative theory, calculated in order of $\alpha_{EM}$. In the vicinity of a very high electromagnetic field, the ability to perfom perturbative predictions in QED disappear, and alternative calculations-methods have to be employed. 

This behavior is expected above a critical field value, which is also known as the Schwinger-limit, and that was first computed in 1951~\cite{Schwinger:1951nm}:
\begin{equation}
\ecrit=\frac{m^{2}_{e}c^{3}}{\hbar{}e}\approx{}1.3\times{}10^{18}~V/m,
\end{equation}

where $m_{e}$ is the electron mass.

Over this limit, electron and positron pairs are created through the Breit-Wheeler~\cite{Breit:1934zz} process; in Compton interactions the kinematic edges that appear at specific energies (also known as Compton edges), are shifted to lower values. This regime can be called Strong Field QED (SFQED)~\cite{Heisenberg:1935qt}. 

SFQED can be accessed in LUXE~\cite{Abramowicz:2021zja} for the first time by interacting the $16.5~$GeV electron beam from the European XFEL with a high-intensity multi-Terawatt Laser system. The strong electromagnetic-field is obtained by the Laser system and is amplified by the Lorentz boost of the electrons. 
In such an environment electrons and positrons pairs are mostly created through a two step process, first a high-energy electron interacts with multiple low-energy photons from the Laser, noted $\gamma_{L}$, giving a high-energy photon, also known as Compton photon and noted $\gamma_{C}$ and an electron. This reaction is also known as non-linear Compton interaction:
\begin{equation}
e^{-}+n\gamma_{L}\rightarrow{}e^{-}+\gamma_{C}.
\end{equation}

In a second step, the Compton photon interact with multiple low-energy photons from the Laser, giving the electron positron pair. This reaction is also known as non-linear Breit-Wheeler pair production:
\begin{equation}
\gamma_{C}+n\gamma_{L}\rightarrow{}e^{+}+e^{-}.
\end{equation}

LUXE also plans to run in a mode where the Laser will interact directly with high-energy photons, that will either be produced through Bremsstrahlung or through Inverse Compton Scattering (ICS). This $\gamma$--Laser running mode is interesting to study, since it is only sensitive to the non-linear Breit-Wheeler pair production.

Two quantities are introduced to characterise these interactions:\\
the classical non-linearity parameter ($\xi$), which measure the electron--Laser coupling and the Laser intensity
\begin{equation}
\xi=\frac{m_e}{\omegaL} \frac{\enLaser}{\ecrit},
\end{equation}

where $\omegaL$ is the Laser frequency and $\enLaser$ is the instantaneous Laser field strength;

the Quantum non-linearity parameter, whose squared measures the fraction of Laser energy transferred to electron beam
\begin{equation}
\chi_{i}=\frac{\epsilon_{i}}{m_e}\frac{\enLaser}{\epsilon_{crit}}(1+\beta\cos\theta),
\end{equation}

where subscripts i are used to denote particle type (“e" for an electron parameter and “$\gamma$" for a photon parameter), $\epsilon_{i}$ is the particle energy, and $\theta$ is the
 collision angle of the particle with the Laser pulse such that $\theta= 0$ is “head-on”. $\hbar=c=1$ has been used, $\beta=1$ for photons and $\beta\approx 1$ for electrons. 
 
In the perturbative regime, $\xi<<1$ the pair production rate is proportional to a power law. In the non-perturbative regime, $\xi>>1$ the rate is not anymore proportional to a power law. 

Measuring with precision the positron appearance rate and the Compton edges positions at LUXE will allow to study in great detail the characteristics of SFQED. 

\begin{figure}[ht]
\centering
\includegraphics[width=0.85\linewidth]{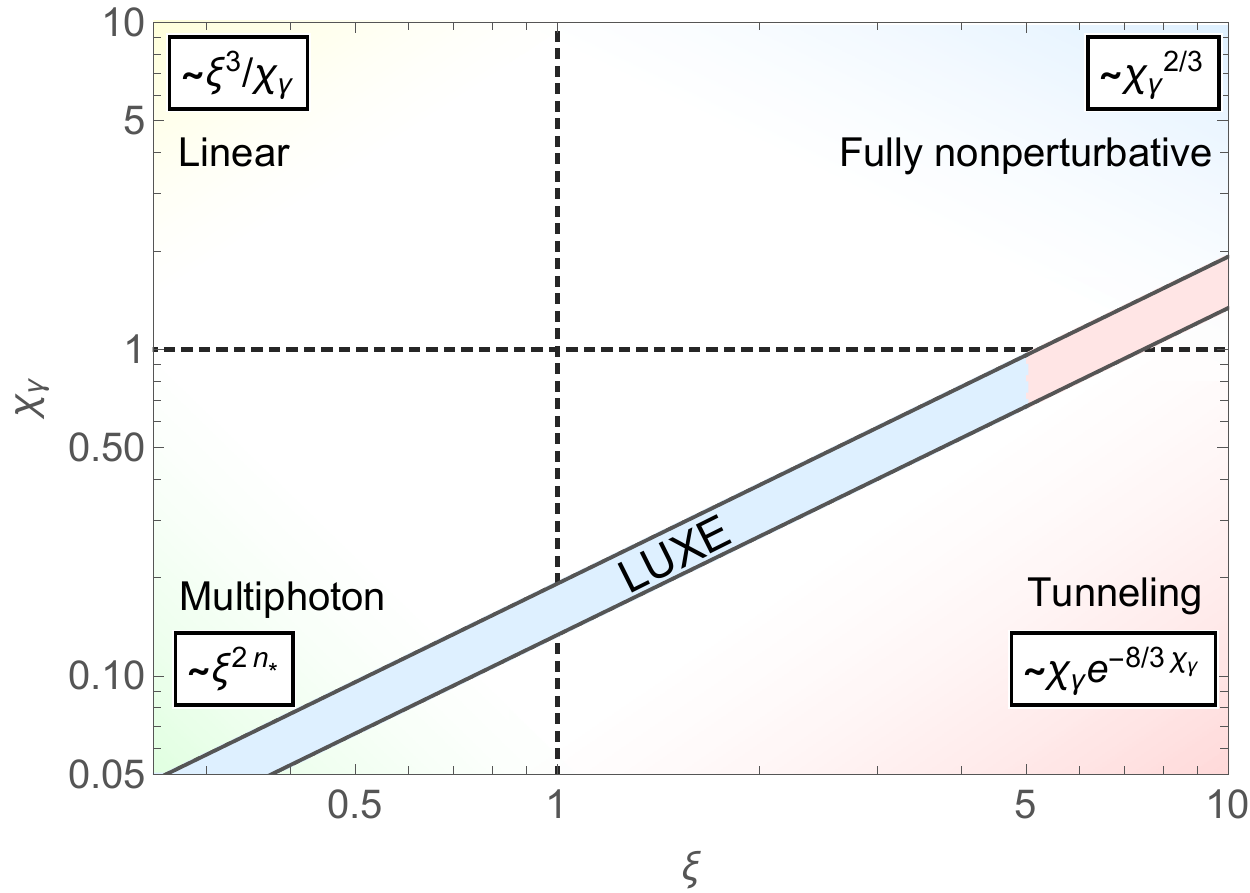}\
\caption{$\xi$ vs $\chi_\gamma$ plane that can be  probed by LUXE.}\label{fig:theoryPlane}
\end{figure}

Figure~\ref{fig:theoryPlane} shows the phase space that will be probed by LUXE in the $\xi$ vs $\chi_\gamma$ plane and compared to the asymptotic scaling of the Breit–Wheeler predictions.

\section{The European XFEL}

The European XFEL (XFEL.EU)~\cite{XFELTDR} is an underground 3.4~km long scientific instrument, running since 2017, that generates ultra-intense X-ray flashes that are distributed to photon science experiments located in Schenefeld. 

\begin{figure}[ht]
\centering
\includegraphics[width=0.85\linewidth]{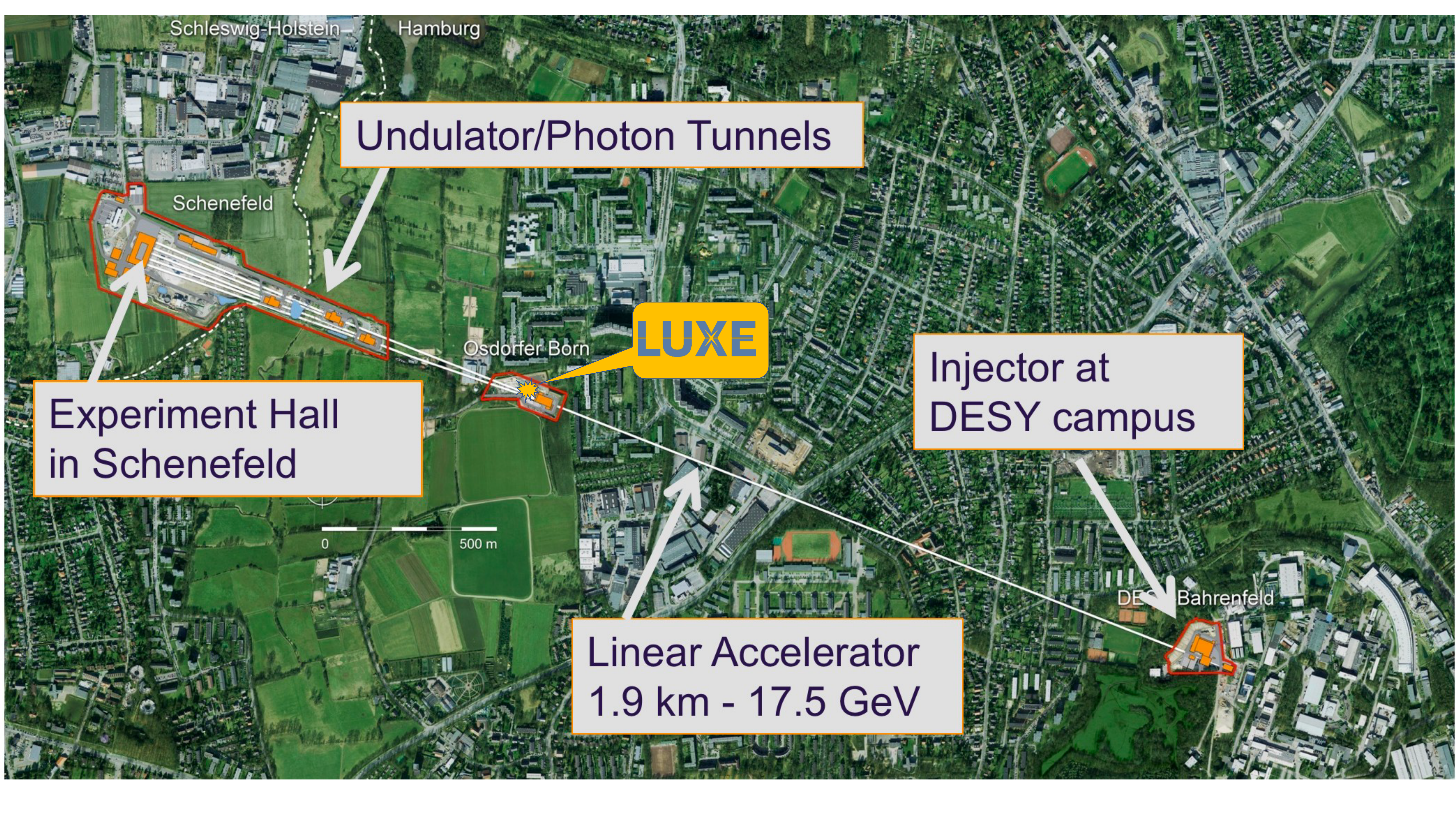}\
\caption{Aerial view of the European XFEL accelerator complex, with the LUXE location highlighted.}\label{fig:EUXFELTop}
\end{figure}

As can be seen in Figure~\ref{fig:EUXFELTop}, the instrument contains a 1.9~km long linear electron accelerator that starts from the main DESY campus. The electrons are distributed to several beam-lines containing undulators that allow to generate the X-ray light thanks to self-amplified spontaneous emission mechanism. 

In the linear accelerator, 2700 electron bunches are created at 10~Hz. Electrons can be accelerated up to $17.5~$GeV, but the current highest value being used in the complex is $16.5~$GeV. The bunch current expected to be used in LUXE is 0.25~nC which corresponds to an occupancy of about $1.5\times{}10^{9}$ electrons per bunches.

LUXE is planned to be installed at the end of the linear accelerator in a non-instrumented tunnel of the XS1 underground building, which was built to allow an extension of the XFEL.EU scheduled to take place after 2029.

In order to extract the electron bunches and bring them to the experiment, it is planned to create a new extraction and transfer line. The TD20 line~\cite{beamlinecdr} has been designed to reuse current XFEL.EU magnets and control systems, and will contain a new kicker magnet that will allow to extract one electron bunch in about $1\mu$s at the end of the bunch train in order to minimise the impact on photon science. 

This will allow to take about 10 Hz of electron data. At the LUXE Interaction Point (IP), it is currently foreseen to focus the electron beam in the transverse plan at $5~\mu$m. The bunch length is of the order of $100~f$s. 

\begin{figure}[ht]
\centering
\includegraphics[width=0.85\linewidth]{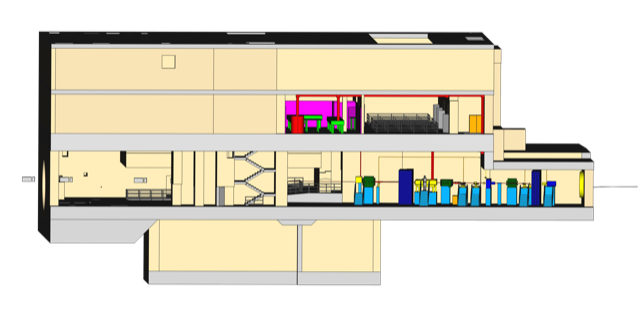}\
\caption{CAD view of the XS1 underground building with LUXE integrated.}\label{fig:CADLUXE}
\end{figure}

Figure~\ref{fig:CADLUXE} shows LUXE being integrated in the current XS1 infrastructure. Since the underground area where the experiment will be installed is not usually accessible during a regular period of data-taking, the installation planing and the access routine of the experiment have been designed to account for it. The experiment is currently planed to be installed during an exceptional 6 months shutdown of the XFEL.EU that will happen in 2024.

\section{High-intensity Laser}

In addition to the XFEL.EU electron beam, LUXE will uses a high-intensity Laser to operate. Since the ground floor of XS1 is not accessible during regular data-taking, the Laser will be installed one floor above, as it can be seen in Figure~\ref{fig:CADLUXE}. 

The Laser used in LUXE employs the Chirped Pulse Amplifcation technique~\cite{strickland1985compression}. This technique  was developed at the end of 1980s and was awarded the Nobel price in 2018 for its breakthrough impact on Laser physics.

LUXE will operate with a Titane--Saphire Laser that peaks at 800 nm wavelength. The experiment will be conducted in two different phases. In phase-0, a custom 40~TW Laser system developed in Jena (JETI40) will be used, while in phase-1 a commercial 350~TW Laser system will be bought. 

High intensity at the IP will be obtained by strongly focusing the beam in both time and space.  Table~\ref{fig:LaserParameters} summarizes the parameters of the Laser used and the expected values of $\xi$ and $\chi$ that will be reached by LUXE. 

The crossing angle between the electron and the photon beam is expected to be about $18^{\circ}$. The Laser pulse will be 30~fs long, and the repetition rate will be 1~Hz. This will leave 9~Hz of electron only data for background studies.

An innovative research and development program for the Laser diagnostics and control system is currently underway since it is expected to reach about 5\% uncertainty on Laser energy measurement and 1\% shot to shot uncertainty.

\begin{table}[ht]
\begin{center}

\begin{tabular}{|l|c|c|c|}
\hline
\textbf{Parameter} & \multicolumn{2}{c|}{Phase 0} & Phase 1 \\
\hline
\textbf{Laser Power [TW]} & \multicolumn{2}{c|}{40} & 350 \\
\hline
\textbf{Laser energy after compression [J]}  & \multicolumn{2}{c|}{1.2} & 10 \\
\hline

   \textbf{Percentage of Laser in focus [\%]}  & \multicolumn{3}{c|}{50}\\   
   \hline
   
    \textbf{Laser focal spot size $w_0$ [$\mu$m] } & $>8$  &  $>3$  &  $>3$\\  
    \hline   
  
   \textbf{Peak intensity in focus [$\times10^{19}$ ~Wcm$^{-2}$]}  &  $1.9$ & $13.3$ & $120$ \\  
    \hline
    
        \textbf{Peak intensity parameter $\xi$}  &  3.0 &  7.9 &  23.6\\   
        \hline   
    \textbf{Peak quantum parameter $\chi$ for $E_e=16.5$~GeV}  &  0.56 &  1.50 &  4.45\\  
           \hline   

\end{tabular}
\caption{Laser parameters user in the different LUXE running phases.}\label{fig:LaserParameters}   
\end{center}

\end{table}

\section{Experiment}

\begin{figure}[ht]
\centering
\includegraphics[width=0.85\linewidth]{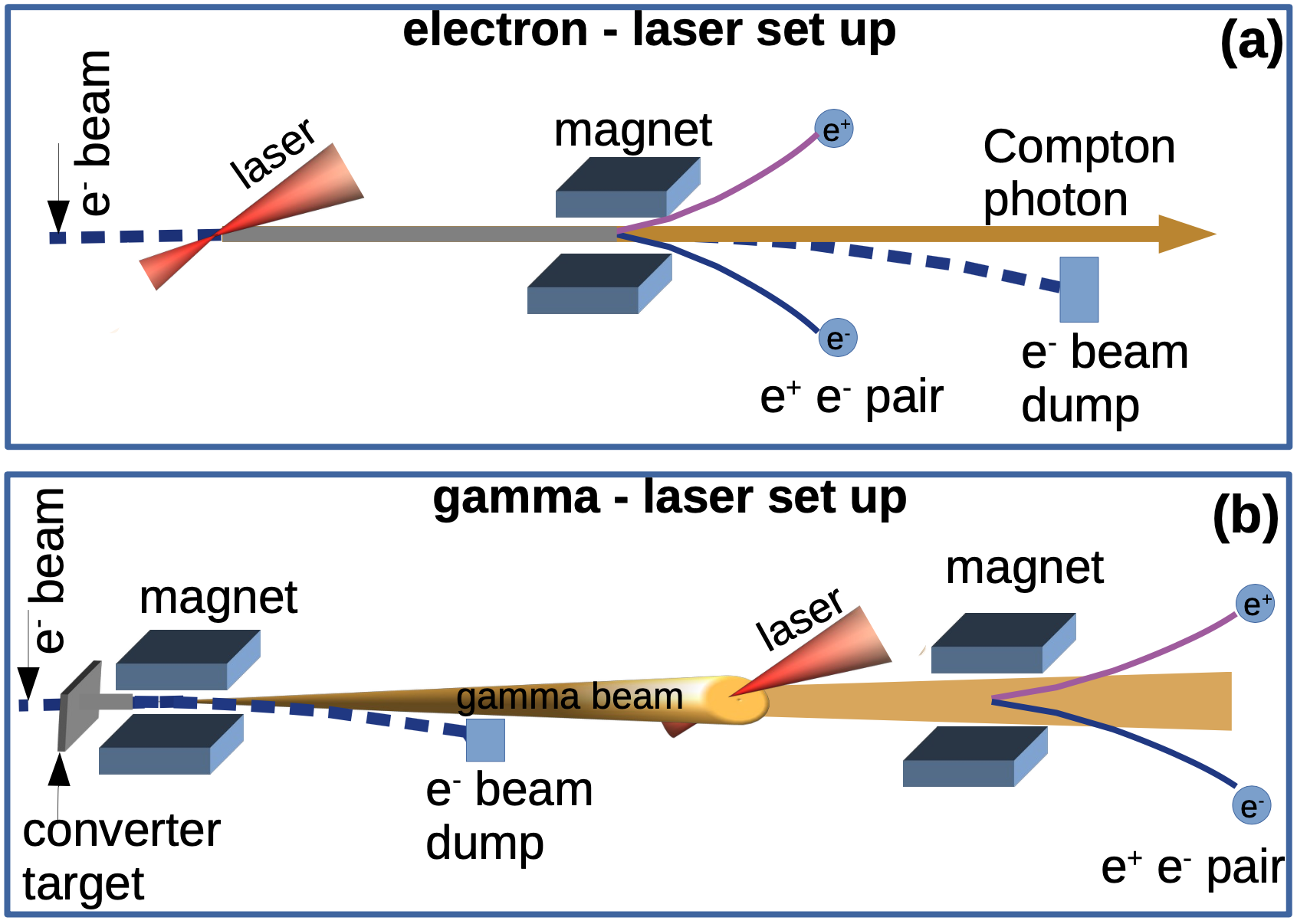}\
\caption{LUXE Running mode (a) $e$--Laser, (b) $\gamma$--Laser.}\label{fig:interactionSketch}
\end{figure}

LUXE will operate in two different data-taking modes, which are depicted in Figure~\ref{fig:interactionSketch}. 

In the electron--Laser setup, shown in Figure~\ref{fig:interactionSketch}~(a), the XFEL.EU electron beam interacts with the Laser, allowing to create Compton photons and consequently electron-positron pairs. These pairs will be separated and analysed using a dipole spectrometer magnet, which is also used to dump the remaining XFEL.EU electron bunch. The spectrometer has been designed with instruments capable of distinguishing the expected signal from the backgrounds. 

\begin{figure}[ht]
\centering
\includegraphics[width=0.85\linewidth]{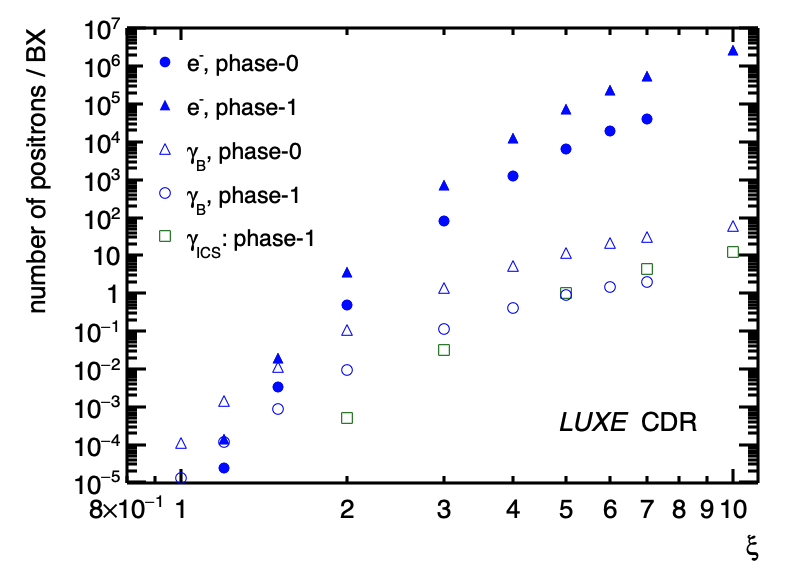}\
\caption{Expected positron rate for the different LUXE running modes and phases.}\label{fig:eLaser_rates}
\end{figure}

Figure~\ref{fig:eLaser_rates} shows the rate of positrons expected for the different running modes and phases of LUXE. As can be seen in the electron--Laser mode this rate varies from about $10^{-4}$ to $10^{6}$ positrons created per XFEL.EU bunch crossing. These numbers have to be compared to the expected number of background particles. On the electron side of the spectrometer one expect up to $10^{9}$ particles per bunch crossing, while on the positron side of the spectrometer one would expect up to $10^3$ particles per bunch crossing. These numbers motivate the usage of radiation-hard detectors (Cherenkov, Scintillating screen) on the electron side of the spectrometer, and precision detectors (Tracker, Electromagnetic Calorimeter) on the positron side.

In the $\gamma$--Laser setup, as seen in  Figure~\ref{fig:interactionSketch}~(b), GeV photons are created upstream of the IP using a Bremsstrahlung target or using another Laser beam for the ICS mode. The XFEL.EU beam is then dumped using a spectrometer magnet. The detectors located after the IP are now protected from most of the background coming from the dump using a large shielding. Thanks to this the amount of expected background on both side of the spectrometer now drops to about $10$--$100$ particles per bunch crossing. As seen in Figure~\ref{fig:eLaser_rates}, the number of signal events per bunch crossing is now expected to be of the order of $10^{-5}$ to $10$ per bunch crossing such that precision detectors (Tracker, Electromagnetic Calorimeter) are used on both side of the spectrometer.

Other detectors systems are placed along the experiment to measure the flux of photons produced at the IP or at the Bremsstrahlung target.

An extra detector will be placed on the final wall of the experimental hall to detect hypothetical Beyond the Standard Model particle, such as Axion Like Particles (ALPS), produced in the dump. Preliminary studies have shown that one year of LUXE data taking will give sensitivity similar, or better, to what is expected from dedicated experiments conducted with the full High Luminosity LHC statistics~\cite{Bai:2021dgm}.

\section{Conclusion}

LUXE is a well thought-through experiment that plans to study in detail QED in the non-perturbative regime. 

It will work with established detector technology to cope with the challenging range of particles rates to measure. 

From the Laser side, an innovative diagnostic and control system development is ongoing to precisely control the energy and reproducibility of each shot.

The experiment will be installed in the XFEL.EU complex by 2024 during the only currently foreseen long shutdown of the facility. 

\section{Acknowledgements}

This work was in part funded by the Deutsche Forschungsgemeinschaft under Germany’s Excellence Strategy – EXC 2121 “Quantum Universe" – 390833306 and the German-Israel Foundation (GIF) under grant number 1492. It has benefited from computing services provided by the GermanNational Analysis Facility (NAF).

\bibliographystyle{JHEP}
\bibliography{Biblio}

\end{document}